# Tip-leakage-flow excited unsteadiness and associated control

Yabin Liu(刘亚斌),[1, a)] Zhong-Nan Wang(王中南)*,[2] Lei Tan(谭磊),[3] Paul Gary Tucker,[4] and Felix M. Möller[5]
[1)]School of Engineering, University of Edinburgh, Edinburgh, EH9 3BF, UK.
[2)]College of Engineering and Physical Sciences, University of Birmingham, Birmingham B15 2TT, UK.
[3)]State Key Laboratory of Hydroscience and Engineering, Department of Energy and Power Engineering, Tsinghua University, Beijing 100084, China.
[4)]Department of Engineering, University of Cambridge, Cambridge CB2 1PZ, UK.
[5)]Institute of Test and Simulation for Gas Turbines, Department of Virtual Engine and Numerical Methods, German Aerospace Center (DLR), Linder Höhe, 51147 Cologne, Germany.

(*Electronic mail: z.n.wang@bham.ac.uk; corresponding author)

(Dated: 15 May 2024)

Tip leakage flow in turbomachinery inherently generates intense unsteady features, named self-excited unsteadiness, which significantly affects the operating stability, aerodynamic efficiency and noise but has not been well understood yet. A Zonalised LES (ZLES) is employed for a linear cascade, with wall-modelled Large Eddy Simulation (LES) active only in the tip region. The simulation is well validated with the advantages demonstrated for reducing the computational effort effectively while maintaining an equivalent accuracy in the region of interest. The time-averaged and spatial-spectral characteristics of tip leakage vortex (TLV) structures are systematically investigated. The self-excited unsteady processes of TLV include the tip gap separation, the tip leakage and mainstream interaction, the primary tip leakage vortex (PTLV) wandering motion and the induced separation near end wall. The Spectral Proper Orthogonal Decomposition (SPOD) is used to examine the dominant frequencies and their coherent structures. It is found that these unsteady features change from single high frequency to multiple lower frequencies due to the PTLV breakdown. The SPOD and correlation analyses reveal that the self-excited unsteadiness of PTLV originates initially from unsteady vortex separation in the tip gap and is then fed by the interactions between the tip leakage jet and mainstream. The associated unsteady fluctuations are convected along the tip leakage jet trajectory, causing the wandering motion of PTLV core. Based on the revealed cause, a micro-offset tip design is proposed and validated for effectively suppressing this unsteadiness, and associated turbulence generation and hence pressure fluctuations. This work improves the understanding of tip-leakage-flow dynamics and informs the control of the associated unsteady fluid oscillation and noise.

## I. INTRODUCTION

Turbomachinery plays a significant role in the energy supply, such as turbines, and consumption, such as pumps and compressors, in the modern world. The unsteady flows in turbomachinery result in long-standing challenges to turbomachinery performance, among which tip leakage flow is one of the most complex problems. The inevitable narrow gap between the blade tip and endwall in a rotor/impeller induces tip leakage jet (TLJ) from the blade pressure side (PS) to the blade suction side (SS), as shown in Figure 1. The tip leakage flow develops into complex TLV structures, which is regarded as an important source of unsteadiness and flow loss[1–4] in turbomachinery. The tip leakage flow generates significant unsteadiness that leads to excessive noise and detrimental cavitation that have been widely reported in many types of turbomachinery, such as compressors[5], fans[6] and pumps[7]. Consequently, considerable efforts have been taken to investigate the vortex structures and flow mechanics associated with the tip leakage flow.

The existing literature demonstrates that there are two types of unsteady patterns due to tip leakage flow in turbomachinery: one is caused by the impingement between the PTLV and blades; the other is the unsteady pattern inherently excited by the tip leakage flow. The trajectory of PTLV is significantly influenced by the tip gap size[8,9] and other parameters[10], such as rotating speed, blade number and blade cumber. Therefore, under specific circumstances such as a large tip gap size, a high rotating speed and a large number of blades[10], an intense impingement interaction may occur between the PTLV and blades. When there is such an impingement interaction, a significant vortex oscillation occurs and causes intensified pressure fluctuation in mixed-flow pump[11]. This unsteady oscillation has been widely studied and is regarded to be closely related to stall[12–14] and generates unsteady blade forces, causing flutter in compressors[15].

However, apart from the unsteadiness caused by the PTLV-blade impingement, tip leakage flow inherently presents an unsteady mode, such as vortex wandering[16], which is excited by the tip leakage jet itself. This self-excited unsteadiness is found to be a potential precursor of rotating instability and dominates the measured

---

[a)]Also at State Key Laboratory of Hydroscience and Engineering, Department of Energy and Power Engineering, Tsinghua University, Beijing 100084, China.

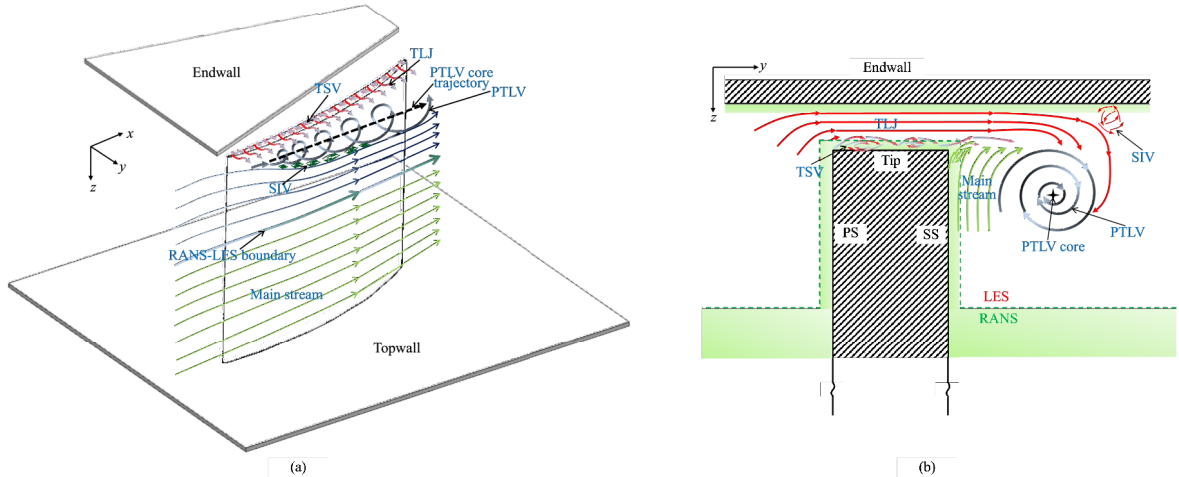

FIG. 1. Schematic of Zonal RANS-LES configuration: (a) 3D view (b) wall-modelling, $y-z$ view.

sound spectrum in a turbo fan[16]. Unfortunately, this unsteadiness is underestimated in most of the previous studies on tip leakage flow, as most of them are based on Reynolds-averaged Navier-Stokes (RANS) simulation, which is based on the Boussinesq hypothesis. This RANS approach relies on time-averaged quantities and tends to smear out the unsteady effects. This limitation is particularly pronounced in the context of multi-scale flows that exhibit high levels of anisotropic turbulence[17,18], which is a distinct feature of the tip leakage flow. Some RANS approaches, like the Reynolds Stress Model (RSM), solve transport equations for individual Reynolds stresses and allow for a more accurate representation of anisotropic turbulence effects[19]. However, due to their time-averaging nature, they still can not fully capture the multi-scale nature of unsteady vortex dynamics. The Particle Image Velocimetry (PIV) measurements[20] demonstrated that the tip leakage flow results in multiple-scale vortex structures, and the large-scale PTLV will break down and collapse into small-scale structures when developing downstream[20,21]. The observation shows that these structures are linked with different unsteady flow features, but how these unsteady flow features are excited and correlated with each other is still unknown. Although several LES works have been conducted to capture the resolved tip leakage vortices[1,22–24], the significant influence of the tip leakage jet has been less clearly identified[25]. The role of the tip leakage jet in generating the self-excited unsteadiness and the evolution process of the related unsteady modes remain unclear. Therefore, the present work focuses on unveiling the generation and propagation dynamics of self-exited unsteadiness.

To understand the fundamental mechanisms, it is crucial to resolve large-scale turbulence motion of the tip leakage flow using LES. Unfortunately, the demanding grid resolution requirement of LES simulation still makes it unaffordable for high-Reynolds number industrial flows. This fact leads to the birth of hybrid RANS-LES models[26], such as the Detached Eddy Simulation (DES) model proposed by Spalart[27], which depends on modelled turbulence scale and mesh size. Subsequently, improvements have been made, such as Delayed Detached Eddy Simulation (DDES) for addressing the grid-induced separation[28] problem and Improved Delayed Detached Eddy Simulation (IDDES) for improving the simulation of wall-bounded turbulent flows[29]. However, users have no control over which region is simulated by LES or RANS, as it depends on the grid size and simulated flows. The 'grey area' between LES and RANS is also a problem, concerning the unphysical delays of critical flow instabilities in sensitive regions due to a slow RANS to LES transition[30]. An alternative is a zonal approach, named ZLES, where experienced users are able to specify the region of LES and RANS based on the prior knowledge of the flow to be simulated[31]. This approach is more suitable for our research problem, i.e. tip leakage flow, so that the expensive LES can be focused only in the tip region.

The present work is performed around a linear compressor cascade. The paper is organized as: First, the ZLES method is introduced and applied to the cascade. The results are compared with the experiment[32,33] and LES results[2] for validation. Second, the mean and unsteady tip leakage flow fields are visualized. Following this, the unsteady flow features of each tip leakage flow structure are investigated to illustrate their energy-spectra features and their correlation with each other by SPOD[34,35] and correlation analysis. Subsequently, based on the revealed physics, we proposed a micro-offset tip design to control the self-excited unsteadiness. Finally, we summarise the most important conclusions and discuss future work.

## II. CASE CONFIGURATION AND NUMERICAL METHODOLOGY

### A. Case configuration

The present research is carried out on the same GE rotor B section blade in the experimental work performed by W. Davenport[32,33]. The case configuration consists of a linear cascade with a fixed stationary topwall and a moving endwall.

The basic geometrical parameters are listed in Table I. The blade chord length $c$ is used as the reference length. The blade pitch $L_y$ is 0.929 $c$, and the span $L_z$ is 1.0 $c$. The coordinate system is adjusted based on the experimental setup, which is described in Figure 19 of Appendix A. The tip gap size is set as 1.65% $c$. The Reynolds number is around $4 \times 10^5$ based on the freestream velocity and chord length. The moving endwall speed $V_{endwall}$ is set to model the relative movement between the endwall and the blade, and it was achieved by a moving belt in the experiment[32,33].

TABLE I. Basic parameters of the investigated cascade

| Parameter | Value |
| --- | --- |
| Blade chord length $c$ | 254 mm |
| Axial chord length $C_a$ | 0.546 $c$ |
| Blade pitch $L_y$ | 0.929 $c$ |
| Blade span $L_z$ | 1.0 $c$ |
| Freestream velocity $U_\infty$ | 26 m/s |
| Moving endwall speed $V_{endwall}$ | 23.6 m/s |
| Tip gap size $\delta$ | 4.2 mm |

### B. Numerical methods

#### 1. Zonal LES

In the present work, a ZLES method is employed to resolve the energetic turbulence of tip leakage flow (Figure 1). To make the simulation affordable, the expensive LES is zonalised in the tip region to achieve high-fidelity prediction of tip leakage flow while the rest region is modelled by RANS.

The SST $k-\omega$ turbulence model is used in the RANS region where the flow field is rarely influenced by the tip leakage flow. The LES mode is activated in the tip region, which is the focus of this research. The dynamic Smagorinsky-Lilly model[36,37] is to calculate sub-grid stresses.

The ZLES is achieved by a blending function $f_b$[38]. The turbulence stresses between RANS and LES are blended in the following way:

$$\tau_{ij}^{Turb} = f_b \tau_{ij}^{RANS} + (1-f_b) \tau_{ij}^{LES} \quad (1)$$

where $\tau_{ij}^{RANS}$ is the RANS portion of the turbulence stresses tensor and $\tau_{ij}^{LES}$ is the LES portion of the modelled turbulence stresses tensor.

Taking advantage of $f_b$, we can generically combine RANS and LES portions, and precisely define the RANS-LES boundary. The RANS and LES configuration is adjusted to obtain the optimal balance between prediction performance and computational effort. This method has been successfully used to predict fan tip and wake flows in a bypass engine configuration[31,39]. This zonal approach is implemented in ANSYS Fluent via User Defined Function by defining $f_b$. The region of interest near the tip is running in LES mode ($f_b$=0), while the rest is in the RANS mode ($f_b$=1). The RANS-LES boundary is determined by a streamline marked in Figure 1(a) separating the tip leakage flow from the rest flow region. Figure 1(b) shows the wall-modelling strategy in LES region. RANS is used to model the inner parts of the boundary layer and reduce the required grid resolution of streamwise $\Delta x^+$ and spanwise $\Delta z^+$ compared to the wall-resolved LES. The RANS layer is applied to where a wall distance $\Delta y^+$<100, which approximately covers the viscous sublayer and buffer layer. The grids in the tip region and testing of the independence of the grid resolution and the RANS-LES boundary are presented in Appendix B.

#### 2. Numerical schemes and boundary conditions

The Central Differencing scheme is employed for spatial discretization, and the Second Order Implicit backwards difference method for time integration. The SIMPLEC scheme is used to decouple pressure and velocity in solving the incompressible flow governing equations. In the transient simulation, the time step size is set as $\Delta t = 8 \times 10^{-6}$s $\approx \frac{1}{1250} \times c/U_\infty$, which guarantees that the Courant number is below 1 in the region of primary interest.

The velocity inlet condition is applied at the inlet boundary, and the outflow condition, where a zero diffusion flux for all flow variables, is applied at the outlet boundary. A periodic boundary condition is adopted on the sidewalls to account for the adjacent passages. The no-slip wall condition is applied on the blade surface and the moving endwall, while a free-slip boundary condition is set on the topwall.

## III. VALIDATION OF NUMERICAL METHOD

Figure 2 shows a good agreement of the time-averaged pressure coefficient $\bar{C}_p$, where $C_p = \frac{p}{\frac{1}{2}\rho U_\infty^2}$, between the present simulation and the experiment[32] at the mid-span. In the experiment[33], Reynolds normal stresses were measured on a line crossing the TLV core on $X = 1.51 C_a$, located in the wake regions and is marked in Figure 20 of Appendix A.

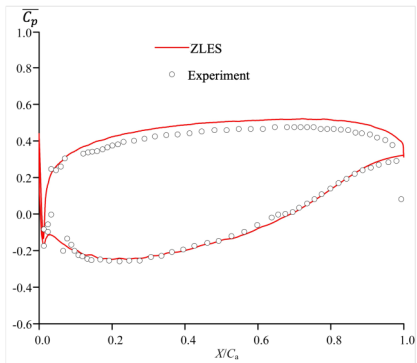

FIG. 2. Comparison of pressure coefficient $C_p$ distribution on middle span between the present simulation and the experiment[32].

To validate the accuracy of the ZLES method, comparison of Reynolds normal stresses has been made between the simulation and the experiment in Figure 3. Moreover, the LES results from You et al.[2] are also plotted as a reference for the validation. Generally, the ZLES method shows an equivalent accuracy to the LES but at a reduced cost due to about six times coarser mesh. Figure 4 shows that the predicted velocity power spectral density agrees well with the experimental measurements[33]. Both the experiment and simulation show the -5/3 slope characteristic in the inertial subrange of turbulence.

## IV. SELF-EXCITED UNSTEADINESS OF TIP LEAKAGE FLOW

### A. Time-averaged flow patterns

This section discusses the time-averaged and spectral characteristics of tip leakage flow. Eight streamwise planes are selected at $x/c$=0.2, 0.3 ... , 0.9 to monitor and visualise the tip leakage flow features. A slice at $z/c$=0.0118 inside the gap, which is normal to the spanwise direction and named Clip-Gap, is extracted to investigate the tip leakage jet dynamics. A slice cutting approximately through the time-averaged vortex core trajectory, named Clip-PTLVC, is set to investigate flow features near the vortex core. Their locations are shown in Figure 20 of Appendix A.

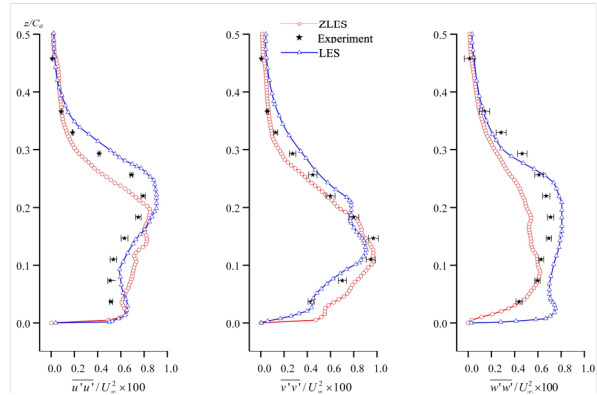

FIG. 3. Comparison of Reynolds Normal Stress along a line crossing the PTLV core on $X = 1.51 C_a$ between the present simulation and the experiment[33].

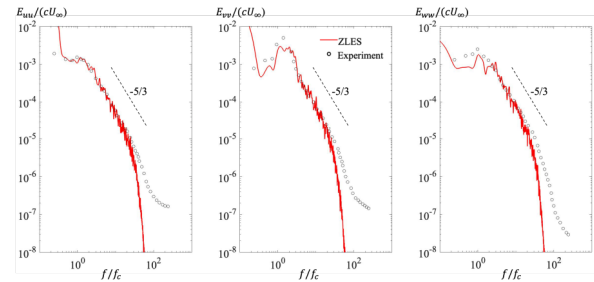

FIG. 4. Comparison of velocity power spectral density at the PTLV core point on $X = 1.51 C_a$ between the present simulation and the experiment[33].

Figure 5 shows the time-averaged streamlines at the tip region from $x/c$=0.3 to 0.6, contoured by the time-averaged streamwise vorticity $\Omega_x$. There are three distinct categories of vortical structures:

1) PTLV: The PTLV is generated by the tip leakage jet and is the dorminant vortical structures in the tip region. It detaches the blade from $x/c = 0.28$, breaks down at $x/c \sim 0.5$, and shows a significant wandering motion. Its unsteady features will be discussed in detail in Sections IV B and IV D.

2) tip separation vortex (TSV): The TSV occurs in the tip gap and initiates near the blade pressure side. Strong shear occurs between them generates K-H rollers, shown by the instantaneous vortex structure defined by $Q$ criterion in Figure 6.

3) secondary induced vortex (SIV): The SIV is induced by the primary tip leakage vortex and rotates in an opposite direction of PTLV. It gradually grows from $x/c$=0.3 and disappears after $x/c$=0.4. This component is not extensively discussed as its influence is limited.

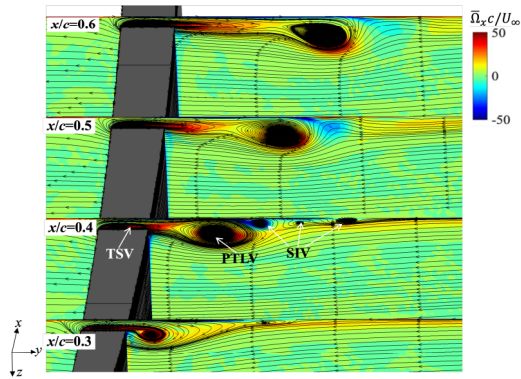

FIG. 5. Time-averaged streamlines and streamwise vorticity on the sections of $x/c$=0.3 to 0.6.

## B. Breakdown of PTLV

To better understand the tip leakage flow features from a 3D and transient perspective, Figure 6 presents the tip leakage flow streamlines that roll up and form the PTLV, which is also illustrated by an iso-surface of $Q$ criterion. The pattern of the streamlines demonstrates the rolling-up process of the PTLV, and the rotation strength of the vortex core decreases when developing downstream, indicated by the condensed streamlines that are entrained into the vortex before the mid-chord and the dispersed streamlines after the mid-chord. A large-scale and coherent vortex structure of PTLV is observed upstream before the mid-chord, and it gradually evolves to multiple small-scale structures downstream after the mid-chord. Therefore, we suspect the PTLV breaks down around the mid-chord location.

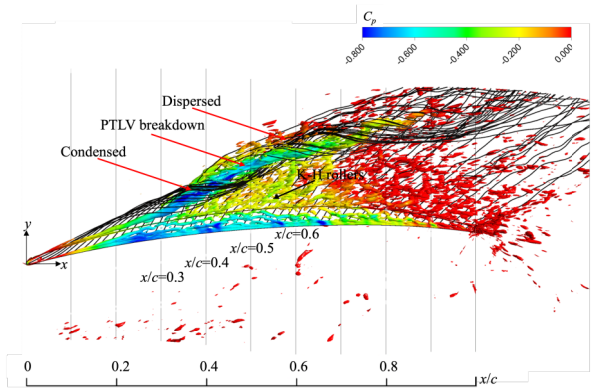

FIG. 6. Instantaneous streamlines of tip leakage flow, with the vortex defined by the iso-surface of $Qc^2/U_\infty^2 = 8 \times 10^2$.

Rossby number $Ro$ is used to quantitatively identify the vortex breakdown behaviour and location. The $Ro$ is a ratio of the axial and circumferential momentum in a vortex, defined as $Ro = U/r\Omega$, where $U$ is the core axial velocity, $\Omega$ is the rotation rate, and $r$ is the vortex core radius[40]. It is a common criterion for vortex breakdown[41]. Based on the time-averaged results, we plotted the distribution of $Ro$ along the PTLV core trajectory in Figure 7a. We found that the $Ro$ at $x/c$=0.5 is close to the critical value of vortex breakdown found by Robinson *et al.*[40]. The decrease of the $Ro$ beyond this critical value reflects the significantly weakened ability of the PTLV to entrain the tip leakage flow from the gap into the vortex core. This change caused by vortex breakdown has a dominant influence on the self-excited unsteadiness that will be discussed in Section IV C.

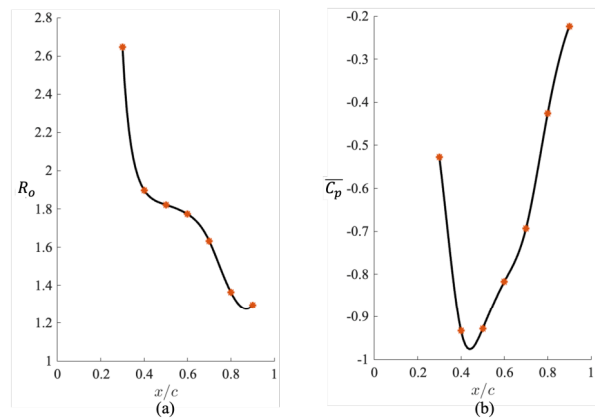

FIG. 7. Rossby number and pressure coefficient along the PTLV core trajectory: (a) $Ro$ (b) $C_p$.

The distribution of time-averaged pressure coefficient $\bar{C}_p$ along the PTLV core trajectory (cf. Figure 7b) further supports the fact that the PTLV breaks down at $x/c \sim 0.5$. $\bar{C}_p$ at the vortex core first dramatically decreases in the early stage but significantly lifts up after $x/c$=0.5. This is caused by the increased static pressure due to the slower fluid rotational motion after the vortex breakdown. The pressure at the PTLV core rapidly rises after the vortex breakdown, which causes a reduced pressure difference between the pressure side and suction side that drives the tip leakage flow from the gap to the vortex core; this may contribute to the distinct difference in unsteady features before and after the PTLV breakdown, as will be shown in Section IV C. Furthermore, the results imply that the pressure in the vortex rapidly increases after vortex breakdown. This is crucial to the control of vortex-induced pressure drop and cavitation.

## C. Spatio-temporal features

Subsequently, we demonstrate the spatial features of turbulent kinetic energy (TKE) on the streamwise sections from $x/c$=0.3 to 0.8. The tip leakage flow region is divided into the TLJ subzone and the PTLV subzone (Figure 8d), as an effort to clarify their distinct features better and identify their mutual correlations in the following analyses. As shown in Figure 8, the TLJ subzone contains the highest level of TKE, and the high TKE extends along the TLJ direction towards the middle of the passage at the downstream stations. The TKE in the PTLV subzone first increases from $x/c$=0.3 to 0.5, then decreases in intensity and diffuses into a larger region, which is related to the breakdown of PTLV at $x/c$ ∼0.5. Meanwhile, a significant TKE distribution near the endwall is observed from $x/c$=0.4, especially at $x/c$=0.4 and $x/c$=0.5, which is corresponding to the SIV, as shown in Figure 5.

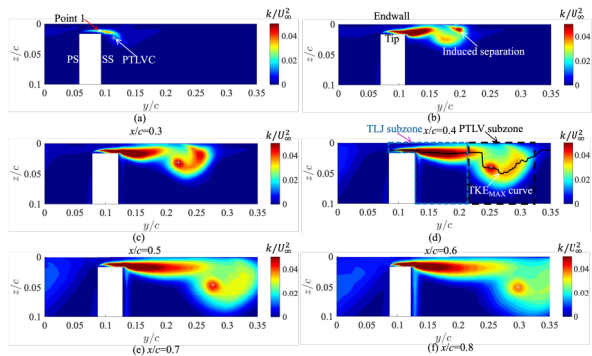

FIG. 8. Distribution of TKE on the streamwise sections of $x/c$=: (a) 0.3 (b) 0.4 (c) 0.5 (d) 0.6 (e) 0.7 (f) 0.8.

To further understand the spectral characteristics of flow unsteadiness along the tip leakage flow propagation, the *PSD* (Power Spectral Density) of velocity is discussed along the TLJ direction and the PTLV core (PTLVC) trajectory, respectively.

Along the trajectory of maximum TKE location at each streamwise section, present in Figure 8(d) for $x/c$=0.6, the spectral features of the tip leakage flow are investigated by plotting the *PSD* of velocity. The horizontal axis of *PSD* map in Figure 9 is the pitch distance, and additional 3 vertical lines are added to show the tip PS, SS and PTLV core locations. The vertical of *PSD* map is the normalised frequency, and the contours show the nondimensionalised *PSD* values.

As shown in Figure 9, the *PSD* map at $x/c$=0.3 shows a single frequency at $f_{S1}$ when the PTLV just detaches from the blade tip. This frequency is limited in the tip gap region and found to be the same as the TSV separation frequency, which will be discussed in Section IV E. When the tip leakage flow develops downstream to $x/c$=0.4 and $x/c$=0.5, the same frequency $f_{S1}$ appears in the PTLV subzone, especially in the PTLV core and induced endwall separated region. It indicates that this unsteady perturbation at $f_{S1}$ could propagate from the upstream tip gap along the TLJ direction to the downstream PTLV core. Further downstream, the PTLV has already broken down and introduced several low frequencies, such as $f_{S2}$, $f_{S3}$ in the TLJ and PTLV subzones. After $x/c$=0.6, the *PSD* level at $f_{S1}$ on the PTLV core significantly decreases, and a group of frequencies centred at $f_{IMP}$ become dominant at $x/c$=0.7 and $x/c$=0.8 where the PTLV gets close to the adjacent blade. This is caused by the impingement of the PTLV on the adjacent blade, which has been studied by He et al.[15].

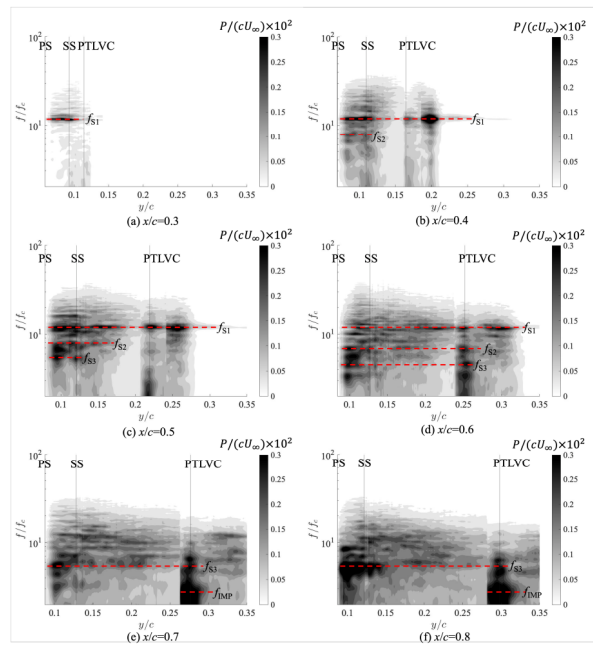

FIG. 9. *PSD* Map of velocity, where $P = \sqrt{(E_{uu}^2 + E_{vv}^2 + E_{ww}^2)}$, on the streamwise sections of $x/c$=: (a) 0.3 (b) 0.4 (c) 0.5 (d) 0.6 (e) 0.7 (f) 0.8. The horizontal axis shows the spatial distribution of $P$ along the $TKE_{MAX}$ curve indicated in Figure 8d. The vertical axis shows the spectral distribution of $P$. PTLVC denotes the mean PTLV core location.

The above discussions demonstrate the statistical characteristics of tip leakage flow, suggesting distinct spectral features along the PTLV. So far, the underlying physics of the self-excited unsteadiness, excluding the unsteadiness caused by vortex-blade impingement[15], remains unveiled. The self-excited unsteadiness is dominant at the early stage of the PTLV evolution and has a significant contribution to the flow losses[25] and noise generation[42]. Therefore, the following section will focus on the underlying physics of these unsteady flow processes, their correlation and the cause of the unsteadiness.

### D. Wandering motion of PTLV

In the PTLV subzone, the most distinct unsteady feature is the PTLV wandering motion after it detaches from the blade, which is corresponding to the observed frequencies in Section IV C. The PTLV wandering motion is shown in Figure 24 (Multimedia available online) in Appendix C, by presenting the instantaneous vortex core locations and the vector field on the streamwise section of $x/c$=0.6.

Figure 10(a) presents a collection of instantaneous positions of the PTLV core. The vortex core is wandering in the $y-z$ plane, where $y$ is the pitchwise direction and $z$ is the spanwise direction. The wandering motion amplitude in the $y$ or $z$ directions is defined as:

$$d_w = \sqrt{\frac{1}{N}\sum_{i=1}^{N}(d_i - \overline{d})^2}, \quad \overline{d} = \frac{1}{N}\sum_{i=1}^{N} d_i \quad (2)$$

where $d_i$ denotes the instantaneous $y$ or $z$ at the snapshot $i$.

Figure 10(b) shows the growth of PTLV wandering amplitude along the streamwise direction. The wandering amplitude increases almost linearly as the vortex travels downstream. Due to the end wall restriction, the amplitude of PTLV is smaller in the spanwise direction than in the pitchwise direction.

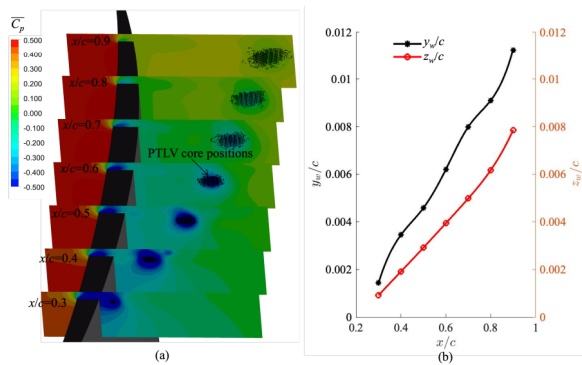

FIG. 10. Vortex core wandering motion along the PTLV core trajectory: (a) Instantaneous core positions (b) PTLV core wandering amplitude: $y_w/c$ and $z_w/c$.

The unsteady features of PTLV are examined by performing SPOD analysis in the spectral domain on the section Clip-PTLVC, which roughly crosses the mean PTLV core trajectory (cf. Figure 1 and Figure 20). The turbulence coherent structures are extracted by SPOD at three characteristic frequencies $f_{S1} = 11.85 f_c$, $f_{S2} = 8.65 f_c$ and $f_{S3} = 5.47 f_c$, where $f_c = U_\infty/c$. As shown in Figure 11, the first SPOD mode at $f_{S1}$ shows a wave-packet pattern and starts when the PTLV detaches the blade. Before the PTLV breakdown, the SPOD mode at $f_{S1}$ is amplified along the PTLV core trajectory; after the PTLV breakdown, it gradually attenuates when developing downstream. With regard to the first SPOD mode at $f_{S2}$ and $f_{S3}$, they both start after the PTLV breakdown, with increasing energy when developing downstream.

The above results suggest that the PTLV vortex breakdown directly influences the spectral feature of the PTLV wandering motion. It changes from a single high-frequency wandering pattern to a more chaotic and multiple low-frequency oscillation.

### E. Vortex separation and shear interaction near the tip

To understand the underlying flow physics in the TLJ subzone at different stages, we demonstrate in Figure 12 the instantaneous vector field on the three streamwise sections of: (a) $x/c$=0.2, before the PTLV detaches the blade; (b) $x/c$=0.4, after the PTLV detaches the blade and before the PLTV breaks down; (c) $x/c$=0.6, after the PTLV breaks down. The flow pattern is almost steady when the PTLV just forms and is attached to the blade surface (at $x/c$=0.2). After the PTLV detaches from the blade, there is an intense unsteady separation inside the tip gap and a significant shear occurs between the tip leakage jet and mainstream outside the gap. This results in significant flow loss and unsteady fluid oscillation inside and outside the tip gap.

Furthermore, the unsteady TSV separation frequency is about $f_{S1}$ at $x/c$=0.4 (cf. Figure 12b, e, h), which is before the PTLV breakdown, and about $f_{S3}$ at $x/c$=0.6 (cf. Figure 12c, f, i), which is after the PTLV breakdown. These results suggest that the spectral features in the TLJ subzone follow the same trend as PTLV, and the instantaneous flow field reflects the unsteady processes corresponding to the identified frequencies.

A further SPOD computation is carried out on the section Clip-Gap, which crosses the TSV separation and TLJ-mainstream shear layer (illustrated in Figure 20). It is clearly shown in Figure 13 that the first SPOD mode at $f_{S1}$ mainly exists before the PTLV breakdown, while that at $f_{S2}$ and $f_{S3}$ emerge after the PTLV breakdown. These findings imply that the PTLV breakdown has the same impact on the spectral characteristics in the TLJ zone as that in the PTLV zone.

### F. Correlation of the unsteady features between the TLJ and PTLV

The above discussions, especially Figures 11 and 13, imply that the unsteady features in the TLJ subzone and the PTLV subzone are related.

To validate this relatioship, the magnitude-squared coherence function[43] and the cross-correlation function[44] are used to estimate their coherence (in spectral domain) and correlation (in time domain). The magnitude-squared coherence estimate is a function of frequency (Equation (3)), measuring the similarity of two signals in frequency.

$$C_{ab}(f) = \frac{|P_{ab}(f)|^2}{P_{aa}(f)P_{bb}(f)} \quad (3)$$

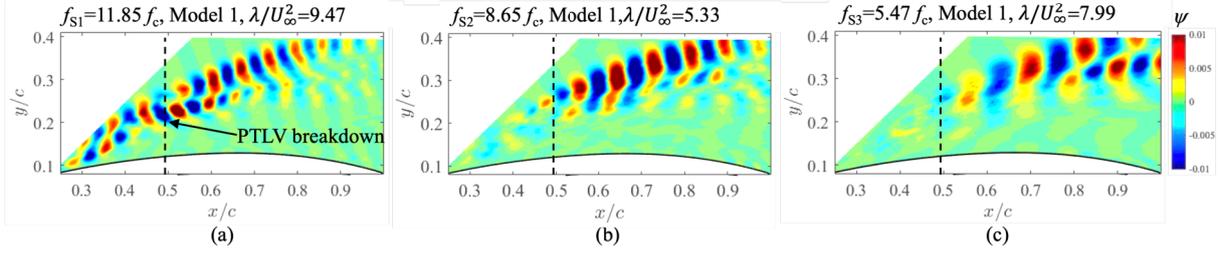

FIG. 11. The first SPOD mode of pitchwise velocity on the monitoring plane Clip-PTLVC at: (a) $f_{S1}$ (b) $f_{S2}$ (c) $f_{S3}$.

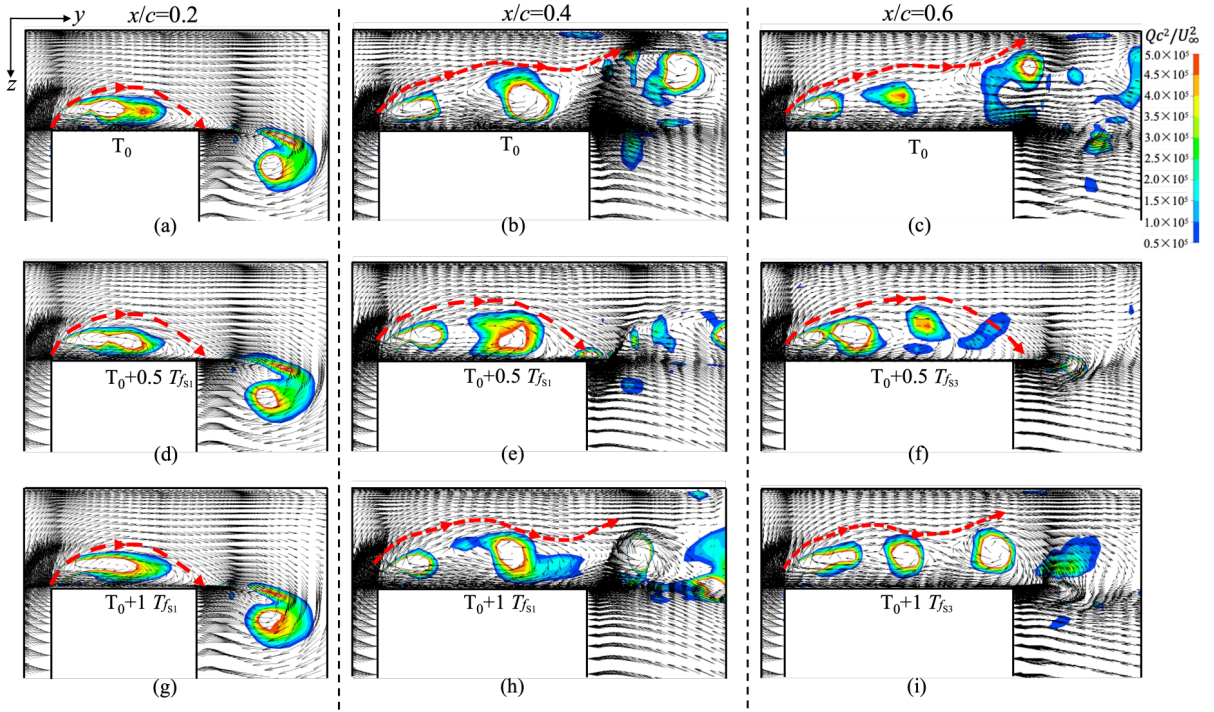

FIG. 12. Instantaneous vector on the streamwise sections of $x/c$=0.2 (a, d, g), $x/c$=0.4 (b, e, h) and $x/c$=0.6 (c, f, i), contoured by the nondimensionalized $Q$ criterion. $T_{f_{S1}}$ denotes the period corresponding to the frequency $f_{S1}$, and $T_{f_{S3}}$ denotes the period corresponding to $f_{S3}$. The relative velocity relative to the moving endwall is applied to the vector.

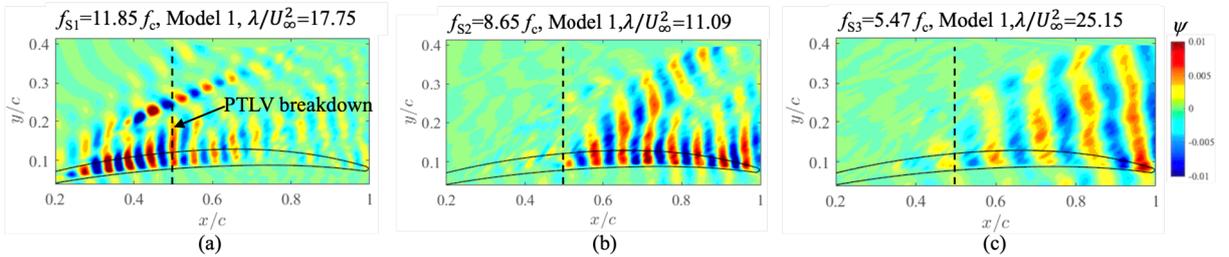

FIG. 13. The first SPOD mode of pitchwise velocity on the monitoring plane Clip-Gap at: (a) $f_{S1}$ (b) $f_{S2}$ (c) $f_{S3}$.

where $P_{aa}(f)$ and $P_{bb}(f)$ are the power spectral densities for signal $a$ and signal $b$, and $P_{ab}(f)$ is the cross power spectral density. $C_{ab}(f)$ denotes the magnitude-squared coherence, measuring the similarity between two signals as a function of the frequency $f$.

The cross-correlation estimate is a function of the time lag (Equation (4) and Equation (5)), measuring the similarity of two signals in time.

$$R_{ab}(\tau) = \int_{-\infty}^{\infty} a(t)b(t+\tau)dt \qquad (4)$$

$$R_{ab}(\tau)_{\text{coeff}} = \frac{R_{ab}(\tau)}{\sqrt{R_{aa}(0)R_{bb}(0)}} \qquad (5)$$

where $R_{ab}(\tau)$ denotes the cross-correlation function, measuring the similarity between two signals ($a$ and $b$) as a function of the time lag ($\tau$) of one relative to the other. $R_{ab}(\tau)_{\text{coeff}}$ is the normalised cross-correlation coefficient using autocorrelations.

By computing the coherence and correlation of the two signals with the above functions, we get the maps of these two coefficients present in Figure 14, respectively. The maps provide a convincing illustration of the spatial-temporal correlation between the tip leakage jet and the PTLV cores. A maximum coherence is observed at the frequency of $f_{S1}$ from $x/c=0.3$ to around $x/c=0.7$, which demonstrates the similarity of the unsteadiness in the tip gap and that at the PTLV cores. The relation between the time lag and the streamwise location is approximately $t_{lag}U_\infty/c = -[(x/c-0.3)+0.012]$, which shows the unsteadiness, excited in the tip gap, convects downstream along the PTLV trajectory. The time lag of $-0.012c/U_\infty$ comes from a time delay between the signal at PTLV core at $x/c=0.3$ and the signal at point 1 in the gap at $x/c=0.3$. This further provides evidence that the PTLV wandering motion originates from the unsteadiness near the tip and propagates downstream along the PTLV core trajectory at the mainstream convection speed $U_\infty$.

## V. FLOW CONTROL BY A MICRO-OFFSET TIP DESIGN

The above results demonstrate that the unsteady separation and shear interaction near the tip is the main source of the self-excited unsteadiness that enhances the turbulent kinetic energy. The LES work performed by You et al.[25] also demonstrated that the significant mean velocity gradients along the spanwise direction lead to the production of turbulent kinetic energy, and the most active turbulent fluctuations are observed near the tip-SS corner. To mitigate this, one of the controlling principles is to decrease the velocity gradient, especially along the spanwise direction, in the near-tip region. Therefore, a micro-offset tip design is correspondingly proposed. As shown in Figure 15, we gradually increase the blade thickness at the suction side in the near tip region, which forces the mainstream direction to be more aligned with the tip leakage jet direction to reduce the spanwise velocity gradient around the tip-SS corner.

In this preliminary test case, the spanwise scope of the tip offset $\tau$ is set as the tip gap size $\delta = 0.0165c$, and the pitchwise scope $\zeta$ is set as 25% of the original tip thickness. The grazing between the TLJ and mainstream at the gap exit is suppressed, and thus the unsteady TSV separation is also significantly mitigated. As a result, the PTLV core wandering is remarkably suppressed. These

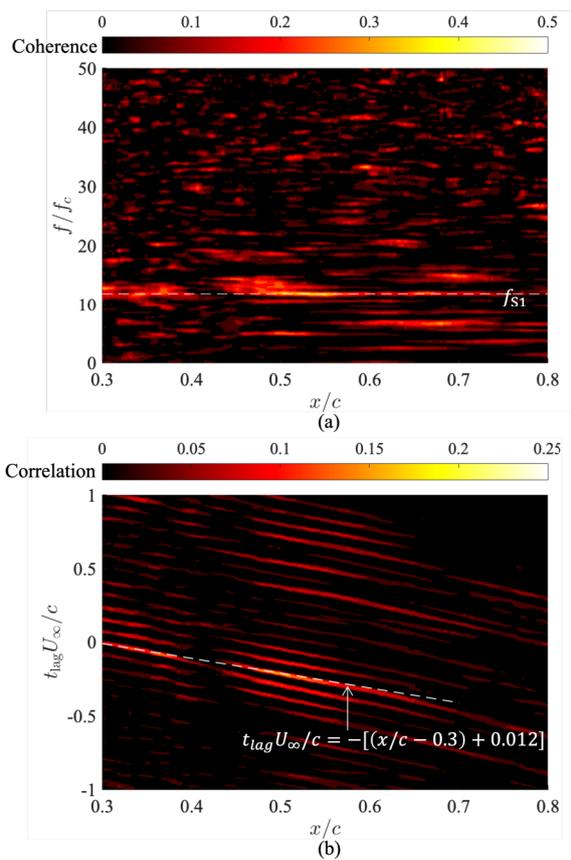

FIG. 14. Coherence and correlation of pitchwise velocity between points 1 in Figure 8a and the approximate PTLV core trajectory on the monitoring plane Clip-PTLVC: (a) magnitude-squared coherence estimate (b) cross-correlation.

In the present work, one signal is taken as the instantaneous pitchwise velocity at point 1 at $x/c=0.3$, as marked in Figure 8a; the other signal is taken as the instantaneous pitchwise velocity along the approximate PTLV core trajectory determined by the $\text{TKE}_{\text{MAX}}$ curve on the monitoring plane Clip-PTLVC indicated in Figure 1 and Figure 20.

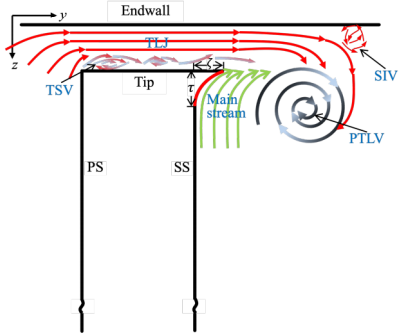

FIG. 15. Schematic of micro-offset tip design.

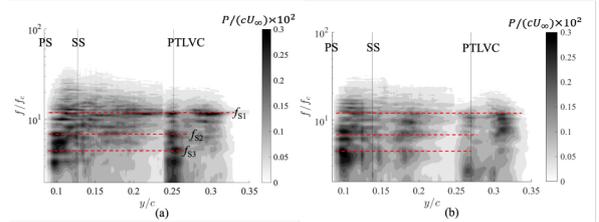

FIG. 17. *PSD* Map of velocity, where $P = \sqrt{(E_{uu}^2 + E_{vv}^2 + E_{ww}^2)}$, on the streamwise sections of $x/c$=0.6: (a) Original blade (b) Micro-offset tip. The horizontal axis shows the spatial distribution of $P$ along the $\text{TKE}_{\text{MAX}}$ curve indicated in Figure 8d. The vertical axis shows the spectral distribution of $P$. PTLVC denotes the mean PTLV core location.

can be observed by comparing the unsteady flow fields of the offset tip with the original one in Figure 25 (Multimedia available online). and Figure 26 (Multimedia available online).

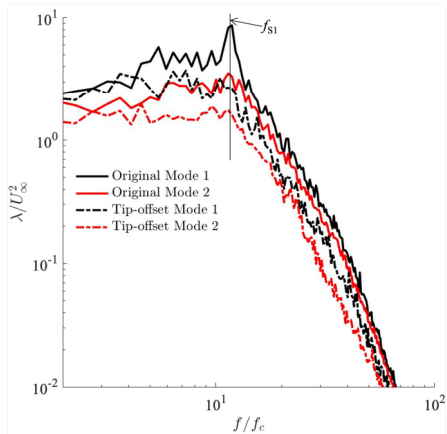

FIG. 16. Comparison of SPOD energy spectra of the tip leakage flow region at $x/c$=0.6 between the original blade and the blade with micro-offset tip.

Figure 16 and Figure 17 show the effect of micro-offset tip on suppressing turbulence induced by tip leakage flow. By taking advantage of micro-offset tip design, the velocity fluctuation in the tip leakage flow region, including the PTLV core, is remarkably reduced. In particular, as shown in Figure 16, the micro-offset tip design can effectively decrease the energy at $f_{S1}$, which is concluded to be the dominant frequency of PTLV core wandering motion. Figure 17 further indicates that the velocity fluctuation of tip separation and TLJ (from the SS to the PTLV core) is also significantly reduced by applying the micro-offset tip design, in addition to the PTLV core itself.

Moreover, as the PTLV core wandering motion is suppressed, the associated pressure fluctuation around the PTLV core is significantly mitigated, as shown in Figure 18. This can help suppress fluid-induced vibration

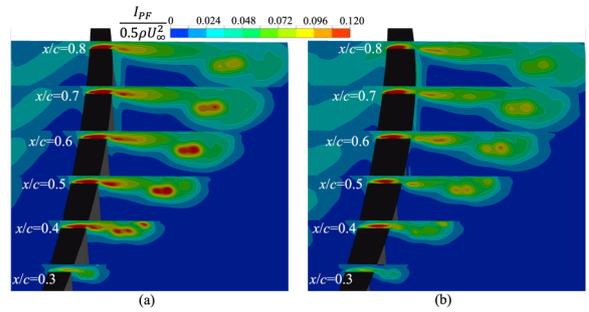

FIG. 18. Effect of micro-offset tip design on suppressing pressure fluctuation: (a) $I_{PF}$ for the original blade (b) $I_{PF}$ for the micro-offset tip design.

and noise in turbomachinery. The pressure fluctuation intensity $I_{PF}$ is defined as:

$$I_{PF} = \sqrt{\frac{1}{N}\sum_{i=1}^{N}(p_i - \overline{p})^2}, \quad \overline{p} = \frac{1}{N}\sum_{i=1}^{N} p_i \qquad (6)$$

where $p_i$ denotes the instantaneous pressure at the snapshot $i$.

## VI. CONCLUSION

The present paper provides insight into the tip-leakage-flow excited unsteadiness around a linear compressor cascade. The following conclusions can be drawn:

(1) As tip leakage flow presents a confined zonal feature, the ZLES approach is a promising way to resolve the interested turbulence structures with locally high fidelity for a reduced computational cost.

(2) The tip leakage flow structures present intense unsteady features. Along the tip leakage jet, there is a significant unsteady separation for the TSV structure inside the tip gap, followed by a TLJ-mainstream shear interaction, then the PTLV core wandering motion and the induced separation near endwall. The PTLV breakdown has a dominant influence on the spectral features of these unsteady flow scenarios, and it leads to a change of unsteady features from a single high-frequency to multiple low-frequencies.

(3) The SPOD computation and the correlation analysis suggest that the investigated unsteadiness is mainly excited by the unsteady TSV separation and TLJ-mainstream shear interaction near the tip. This unsteadiness originates from the tip gap and propagates to the PTLV core. It further causes the PTLV core wandering, which convects downstream along the PTLV core trajectory at the mainstream convection velocity.

(4) Based on the revealed flow physics, a micro-offset tip design is proposed. It can significantly suppress this self-excited unsteadiness, by reducing the interaction between the tip leakage jet and mainstream near the tip suction side. This flow control method can effectively mitigate the PTLV wandering motion, resulting in a remarkable attenuation of the associated turbulence generation and pressure fluctuation.

The above findings suggest that particular attention should be paid to suppressing the unsteady shear interaction near the tip suction side. We can decrease the TLJ-mainstream interaction by changing the tip leakage jet or near-tip mainstream direction through tip shape modification. The identified critical role of PTLV breakdown also suggests that advancing vortex breakdown active/passive fluid injection is likely to be a new route to control the noise and pressure-drop due to a swirling vortex. The revealed flow physics and potential flow control technologies will be beneficial to reduce flow losses, noise and cavitation, which are critical challenges to improving the performance of turbomachinery and aerial/underwater vehicles.

## ACKNOWLEDGMENTS

This work has been supported by the Royal Commission for the Exhibition of 1851 Brunel Fellowship, the UK, ESPRC funded, Supergen Offshore Renewable Energy Hub [EP/Y016297/1], the State Key Laboratory of Hydroscience and Engineering [sklhse-2023-E-03], and the EPSRC IAA Innovation Competition [EPSRC IAA PV111]. In particular, we are very grateful for the one-year international visiting funding from the China Scholarship Council, which enables this joint work.

## AUTHOR DECLARATIONS

**Conflict of Interest**

The authors have no conflicts to disclose.

## DATA AVAILABILITY

The data that support the findings of this study are available from the corresponding author upon reasonable request.

**Appendix A: Case setup**

The original $X$-$Y$-$Z$ coordinate system is different from the velocity coordinate composed of $u$ (streamwise velocity), $v$ (pitchwise velocity), and $w$ (spanwise velocity) in the experiment[32,33]. Therefore, the space coordinate system has been accordingly rotated by 52.5° to be the new $x$-$y$-$z$ coordinate, which is aligned with the $u$-$v$-$w$ coordinate, as shown in Figure 19.

Figure 20 shows the location of the RANS-LES hybrid boundary and the monitoring planes set in the present work.

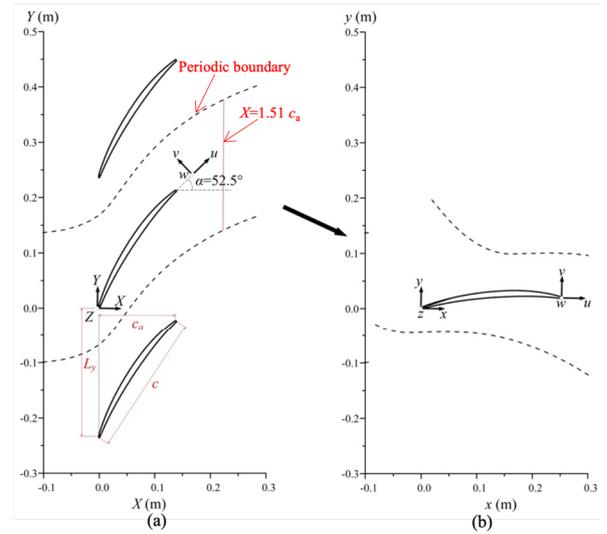

FIG. 19. Adjustment of the coordinate system (top view): (a) Original $X-Y-Z$ coordinate system (b) $x-y-z$ coordinate system aligned with $u-v-w$.

**Appendix B: Validation of numerical methods**

The ZLES method is applied in the present work to save computational effort, so the mesh arrangement is correspondingly adjusted in the LES scope and RANS scope, respectively. Since the tip gap region is in small configuration, the statistical results of two cases with different spanwise blending boundary locations, as shown in Figure 20, are compared. One case is named BD1, and the other case with the RANS-LES boundary shifted by $4\% C$ along $z$ is named as BD2. As shown in Figure 23, a good agreement of the stresses curves between the two cases is observed, which verifies the independence of the RANS-LES blending boundary location.

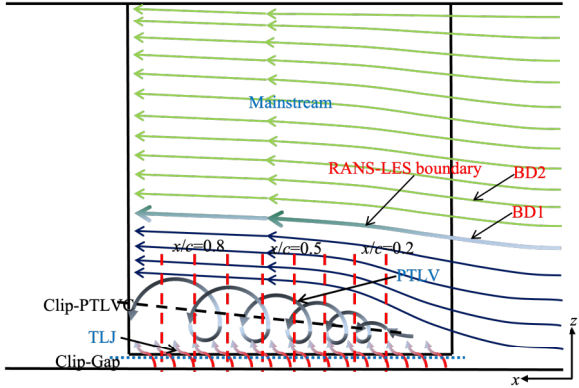

FIG. 20. Schematic of monitoring sections.

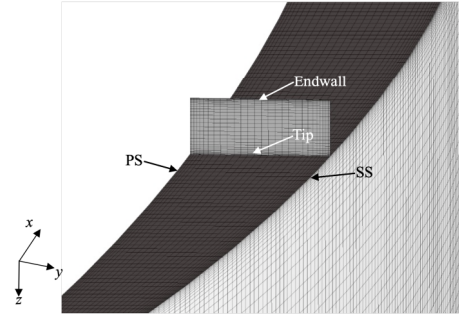

FIG. 21. Mesh on the blade and in the tip gap.

dimension and with bounded wall condition, the grid resolution around the tip gap was locally refined in particular, as shown in Figure 21. The gird resolution in the region of primary interest is controlled in $\Delta x^+ < 60$, $\Delta y^+ < 1$ on blade surface and $\Delta z^+ < 90$. In the region close to the blade tip, the $\Delta z^+$ is refined to 1 2 to accurately capture the flow details near the narrow tip gap. The mesh arrangement in the RANS scope is much coarser, especially for the region far from the RANS-LES boundary, with $\Delta x^+$ and $\Delta z^+$ increased to 50 200. Therefore, the total grid elements number is only 3,531,621 for the 4.2 mm tip gap, with the whole span considered. For the same case, more than 20 million grid elements were used in the LES studies[2,22,23,25], with only half span simulated.

An independence study of grid resolution was also performed to exclude the influence of grid resolution on predicting the turbulence field of the tip leakage flow region. Moreover, the experimental data was measured at $X = 1.51C_a$, and a substantial amount of grid is clustered in the wake region (the region where $X > 1.0Ca$ shown in Figure 19), but this region is actually of slight interest because this study focuses on the tip leakage flow region. Therefore, the prediction performance of the following two cases, named Case 1 and Case 2, is compared. In Case 1, the same grid resolution adopted in the LES studies[2,22,25] is employed in both the tip leakage flow region and the wake region, and these regions are running in LES mode; In Case 2, the grid resolution is the same as described in the above paragraph, while the mesh in the wake region is coarse and running in RANS mode. Furthermore, the Reynolds normal stresses on the three lines from crossing the PTLV core are plotted to compare the prediction performance of Case 1 and Case 2. As shown in Figure 22, the Reynolds normal stress curves obtained by the two cases show a remarkable agreement, which validates that the adopted grid resolution and LES scope in the present work can well predict the turbulence field of the tip leakage flow region.

To further validate the independence of RANS-LES

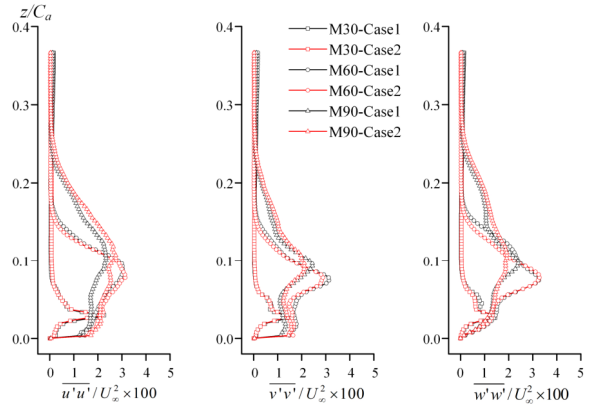

FIG. 22. Comparison of Reynolds Normal Stresses along a line crossing the vortex core between Case 1 and Case 2 with different grids. The line is located on the streamwise sections of $x/c = 0.3$ (M30), $x/c = 0.6$ (M60), and $x/c = 0.9$ (M90), respectively.

### Appendix C: Instantaneous flow field

This section includes the motionless images of the provided multimedia materials.

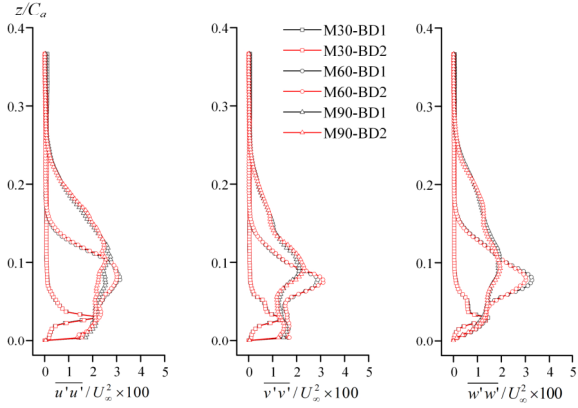

FIG. 23. Comparison of Reynolds Normal Stresses along a line crossing the vortex core between different RANS-LES boundary locations. The line is located on the streamwise sections of $x/c = 0.3$ (M30), $x/c = 0.6$ (M60), and $x/c = 0.9$ (M90), respectively.

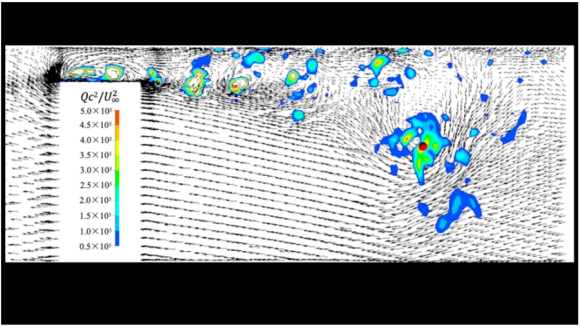

FIG. 24. Instantaneous vector field on the streamwise section of $x/c$=0.6 and vortex core wandering motion. $Q$ denotes the $Q$ criterion. The red point denotes the instantaneous primary tip leakage vortex core location. (Multimedia available online)

## Appendix D: Statistical characteristics of tip leakage flow unsteadiness

Figure 27 shows the computed local SPOD energy spectra from $x/c$=0.3 to $x/c$=0.7, for the TLJ subzone and the PTLV subzone, repsectively. There is increasing low-frequency energy in both subzones when the tip leakage flow develops downstream. Meanwhile, we observe multiple low-frequency peaks from $x/c$=0.5 in the TLJ subzone and from $x/c$=0.6 in the PTLV subzone, which means a streamwise location delay of the spectral features between the TLJ subzone and the PTLV subzone. This supports that the unsteadiness in the TLJ subzone convects to the PTLV subzone.

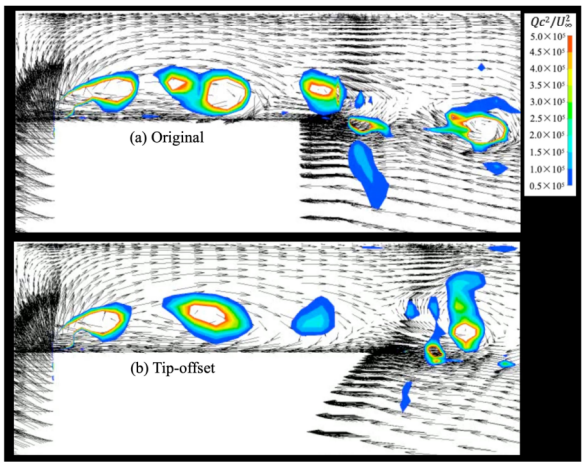

FIG. 25. Comparison of instantaneous vector field near the tip (streamwise section of $x/c$=0.6) between (a) the original case and (b) the micro-offset tip design. $Q$ denotes the $Q$ criterion. (Multimedia available online)

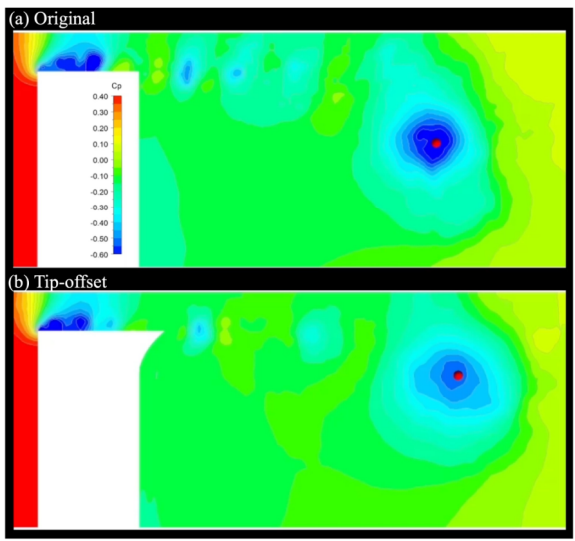

FIG. 26. Comparison of instantaneous pressure-coefficient $C_p$ field (streamwise section of $x/c$=0.6) between (a) the original case and (b) the micro-offset tip design. The red point denotes the instantaneous primary tip leakage vortex core location. (Multimedia available online)

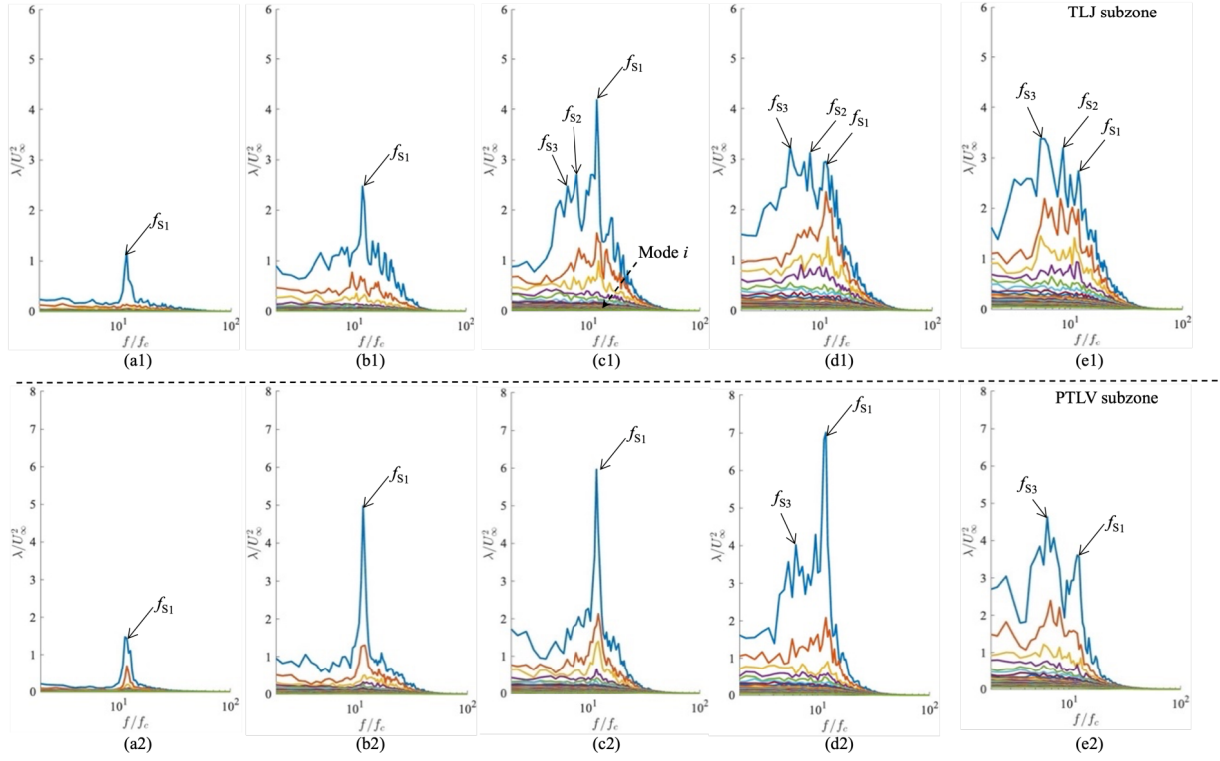

FIG. 27. Local SPOD energy spectra of the TLJ subzone and the PTLV subzone at $x/c=$: (a) 0.3 (b) 0.4 (c) 0.5 (d) 0.6 (e) 0.7.